
\input phyzzx


\catcode`\@=11
\def\binrel@#1{\setbox\z@\hbox{\thinmuskip0mu
 \medmuskip\m@ne mu\thickmuskip\@ne mu$#1\m@th$}%
 \setbox\@ne\hbox{\thinmuskip0mu\medmuskip\m@ne mu\thickmuskip
 \@ne mu${}#1{}\m@th$}%
 \setbox\tw@\hbox{\hskip\wd\@ne\hskip-\wd\z@}}
\def\underset#1\to#2{\binrel@{#2}\ifdim\wd\tw@<\z@
 \mathbin{\mathop{\kern\z@#2}\limits_{#1}}\else\ifdim\wd\tw@>\z@
 \mathrel{\mathop{\kern\z@#2}\limits_{#1}}\else
 {\mathop{\kern\z@#2}\limits_{#1}}{}\fi\fi}
\def\utilde#1{\underset{\widetilde{\ }} \to #1}

\pubnum={192/COSMO$-$17}

\titlepage
\vskip 1cm
\title{\bf{Topological 2-form Gravity in Four Dimensions}}
\vskip 1cm
\author{Hjor Yol Lee,${}^\dagger$
\   Akika Nakamichi \footnote{\star}{e-mail address:
                    anakamic@cc.titech.ac.jp}
\  and \  Tatsuya Ueno \footnote{\diamond}{e-mail address:
                   tatsuya@cc.titech.ac.jp}  }
\vskip 1cm
\address{${\star \, \diamond}$
         Department of Physics, Tokyo Institute of Technology
         \nextline Oh-okayama, Meguro-ku, Tokyo 152, Japan
\nextline
\nextline
        ${}^\dagger$
         Tsukuba Institute of Science and Technology
         \nextline Kamitakatsu, Tsuchiura-shi, Ibaraki 300, Japan  }

\vskip 1 cm
\abstract{ A kind of topological field theory is proposed as a candidate
to describe the global structure of the 2-form Einstein gravity with
or without a cosmological constant.
 Indeed in the former case, we show that a quantum state in the candidate
 gives an exact solution of the Wheeler-DeWitt equation.
 The BRST quantization based on the Batalin-Fradkin-Vilkovisky (BFV)
formalism is carried out for this topological version of the 2-form
Einstein gravity. }

\vfill
\eject

\sequentialequations

\chapter{Introduction }%
 In recent years, much attention has been focused on the global
(topological) aspects of the Einstein gravity.
 For example, three-dimensional gravity possesses non-trivial global
structures both in the classical and quantum levels in spite of the
absence of the local fluctuations in the theory.
 In particular, Witten showed the Einstein action in three dimensions to
be exactly equivalent to a Chern-Simons term for the Poincar{\' e} group
[1].
 Using this correspondence, the theory was shown to be exactly soluble
and further in [2], the possibility of topology-changing processes was
discussed.
 These advances in three-dimensional gravity suggest that one can also
obtain important results by studying the global aspects of the
Einstein gravity in four dimensions.
 However in this case, the presence of the local gravitational waves
prevents us from approaching the global feature of the theory directly.
\par

 In this paper, we propose a kind of topological field theory (TFT) as
a promising candidate which describes the global structure of the
four-dimensional Einstein gravity with or without a cosmological
constant.
 This candidate is obtained by modifying an alternative formulation
of gravity recently developed by Capovilla, Dell, Jacobson and Meson [3].
 This alternative, which we will call the 2-form Einstein gravity, is
expressed using anti-self-dual 2-forms as fundamental variables,
instead of the metric or the tetrad.
 In the presence of a cosmological constant, we demonstrate an explicit
example that an exact global feature in the Einstein gravity is obtained
by examining the TFT in quantum theory.
 Namely  a unique quantum state in the TFT turns out to be one of
exact solutions of quantum constraints in the Einstein gravity,
which include the Wheeler-DeWitt equation.
 Moreover, the BRST quantization for the TFT is performed using the
Hamiltonian formalism introduced by Batalin, Fradkin and Vilkovisky
[4-6].
 This powerful method for constrained systems is expected to be useful
for further investigations of the quantum feature of gravity.
 We get the BRST invariant action in covariant form for the TFT with or
without a cosmological term.
 In particular, in the former case, it is shown that the action is
identical with the one in Witten's topological Yang-Mills theory (TYMT)
for the SU(2) group [7-10].                \par

 The action of the (Euclidean) 2-form Einstein gravity is given in terms
of a symmetric spinor-valued 2-form $\Sigma^{AB}$ and an SU(2) spin
connection 1-form $\omega_{AB}$ in the presence of the cosmological
constant $\Lambda$,
$$
 S = \int {\Sigma^{AB}} \wedge {R_{AB}}
  -{\Lambda \over 24} \  {\Sigma^{AB}} \wedge {\Sigma_{AB}}
  -{\alpha \over 2} \  {\psi_{ABCD}}{\Sigma^{AB}} \wedge {\Sigma^{CD}}
  \ ,                                                         \eqno(1)
$$
where $R_{AB} \equiv d\omega_{AB} + \omega_{AC} \wedge {\omega^C}_B $,
$\, \psi_{ABCD} $ is a totally symmetric Lagrange multiplier field and
$\alpha$ is an arbitrary parameter.
\footnote{{}^{\sharp1}}{Spinor (and primed spinor) indices are raised
and lowered according to the convention in [11].
 The symbol $\wedge$ means the exterior product of forms. }
 In this formulation, the metric field $g_{\mu \nu}$ is defined
in terms of the 2-form $\Sigma^{AB}$ as
$$
g^{1\over 2} g_{\mu \nu}
             = {1 \over 24} \  {\varepsilon^{\alpha \beta \gamma \delta}}
                          \   {\Sigma_{\mu \alpha}^{AB}}
                            \     {\Sigma_{\beta \gamma B}}^C
                              \           {\Sigma_{\delta \nu CA}}\ ,
            \qquad g \equiv det(g_{\mu \nu})\ .
                                                            \eqno(2)
$$
 The constraint equations obtained by varying (1) with respect to
$\psi_{ABCD}$ imply that $\Sigma^{AB}$  is composed of 1-form
$e^{A A^\prime}$ with the SU(2) spinor and `primed' spinor indices [3].
 Namely,
$$
 {\Sigma^{(AB}} \wedge {\Sigma^{CD)}}=0
\iff
 {\Sigma^{AB}} = {\epsilon_{A^\prime B^\prime}}
                        ({e ^{A A^\prime}} \wedge {e^{B B^\prime}})\ .
                                                            \eqno(3)
$$
 Using this result and translating the SU(2) spinor indices into the
SO(4) (the local-Lorentz) indices [11], we find that (1) is identical
with the chiral decomposition of the first order Palatini action in which
 the usual spin connection is replaced by its anti-self-dual part with
respect to the SO(4) indices.
 As is discussed in [12], this chiral action also gives the Einstein
equation.
 Since the action (1) describes general relativity, it is invariant under the
local-Lorentz transformation and diffeomorphism,
$$
  \delta \omega^{AB} = D\theta^{AB}_0 + {\cal L}_\xi \omega^{AB}\ ,
\qquad
  \delta \Sigma^{AB} = [\Sigma, \theta_0]^{AB} + {\cal L}_\xi \Sigma^{AB}
  \ ,                                                       \eqno(4)
$$
where ${\cal L}_\xi$ is the Lie derivative with respect to a vector
field $\xi^\mu$, and the local-Lorentz transformation corresponds to
the SU(2) gauge transformation with a parameter $\theta^{AB}_0$.

\chapter{Topological 2-form gravity}              %
 Let us consider the situation when we drop the last term in (1),
i.e. $\alpha$ = 0.
 {}From now on we use the internal SU(2) indices $ i,j,k,\cdots$ by
expanding ${\Sigma_A}^B = i \Sigma^k {({\sigma_k})_A}^B$  and
${\omega_A}^B$ in Pauli-matrices  $\sigma_k$ for simplicity.
 Then the action (1) with $\alpha$ = 0 becomes
\footnote{{}^{\sharp2}}{ We have omitted an overall factor of 2 in (5).}
$$
  S_{\alpha =0}= \int \Sigma^k \wedge R_k
                   -{\Lambda \over 24} \ \Sigma^k \wedge \Sigma_k \ .
                                                            \eqno(5)
$$
 In this particular case, a new symmetry with a parameter 1-form
$\theta_1^k$ emerges in addition to the local-Lorentz symmetry [3],
\footnote{{}^{\sharp3}}{ We use the notation for the SU(2) indices,
  $F \cdot G \equiv F^i G^i$ and
   $(F \times G)^i \equiv \varepsilon_{ijk} F^j G^k$, where
    $\varepsilon_{ijk}$ is the structure constant of SU(2). \nextline
     $D$ is the spin-covariant exterior derivative, e.g.
       $D\theta_0^k \equiv d\theta_0^k + 2(\omega \times \theta_0)^k$.}
$$
  \delta \omega^k =D\theta_0^k +{\Lambda \over 12}\theta_1^k  \ ,
\qquad
  \delta \Sigma^k =2(\Sigma \times \theta_0 )^k + D\theta_1^k \ .
                                                            \eqno(6)
$$
 Although the theory remains invariant under diffeomorphism just as
in the $\alpha \not= 0$ case of the previous section, there is no need to
 add it to (6) because, modulo the equations of motion derived from (5),
diffeomorphism with the vector fields $\xi^\mu$ can be generated by the
combination of the above local-Lorentz and `new-type' transformations
(6) with
$\theta_0^k
= \xi^\nu \omega_\nu^k$ and $\theta_{1 \mu}^k
= 2\xi^\nu \Sigma_{\nu \mu}^k$.
 With the appearance of the new-type symmetry, the theory turns out
to be on-shell reducible in the sense that the transformation laws (6)
are invariant under
$$
    \delta \theta_0^k = - {\Lambda \over 12} \epsilon_0^k \ ,
  \qquad
    \delta \theta_1^k = D \epsilon_0^k \ ,
                                                            \eqno(7)
$$
if the equations of motion are satisfied.
This means that not all of the parameters in (6) are independent. \par

 For the $\Lambda  \not= 0$ case, the action has such a large symmetry
under $\delta \omega^k \propto \theta_1^k$
($\delta \Sigma^k = D\theta_1^k$) that the spin connection $\omega^k$ can
 be completely gauged away.
 This is very similar to the case of Witten's TYMT for the SU(2) group
[7-10].
 In fact  if one eliminates $\Sigma^k$ from the action (5) by using the
equations of motion, then the effective action for $\omega^k$ reads
$$
   S = {6 \over \Lambda} \int R^k \wedge R_k \ .
                                                            \eqno(8)
$$
 This is the starting action which leads us to the BRST invariant action
of the TYMT in several approaches [8,10].
 Also for the $\Lambda = 0$ case, the action describes a kind of
TFT, called BF theory, which has recently been discussed
[13-15].
 Therefore general relativity, with or without a cosmological constant,
is converted into a TFT if the parameter $\alpha$ is chosen to be zero.
\par

 In our topological model (5), we expect that the modes of the
gravitational wave disappear and there remain no (physical) local
fluctuations.
 To see the change of the physical degrees of freedom and of the
symmetries in this process in detail, let us move into the Hamiltonian
formalism. \par

 In the topological case ($\alpha$ = 0), the action (5) becomes in
canonical form,
$$
   S = \int dt \int d^3x [ \dot \omega_a \cdot B^a
         - \omega _0 \cdot \varphi
            - \Sigma_{a0} \cdot \phi^a ] \ .
                                                            \eqno(9)
$$
 The canonical variables are $\omega_a^k$ and their conjugate momenta
$B^a_k \equiv \varepsilon ^{abc} \Sigma_{bc}^k$,
 which are the spatial components of the spin connection $\omega^k$
and the 2-form $\Sigma^k$.
 Varying (9) with respect to their time components $\omega_0^k$ and
$\Sigma_{a0}^k$, we get two sets of constraints,
$$
   \varphi _k \equiv - D_aB^a_k \approx 0 \ ,
 \qquad
   \phi^a_k \equiv 2 \ ( \varepsilon^{abc}R_{bc}^k
         - {\Lambda \over 12} B^a_k ) \approx 0 \ .
                                                            \eqno(10)
$$
 The Poisson brackets among them are given by
$$
\eqalign{
  & \{ \varphi_i({\bf x}), \varphi_j({\bf y})\} =
    - 2 \ \varepsilon _{ijk} \ \varphi_k({\bf x})
      \ \delta ^3 ({\bf x} - {\bf y}) \ ,
\qquad
  \{ \phi ^a _i({\bf x}), \phi^b_j({\bf y}) \} = 0  \ ,
\cr
  & \{ \varphi_i({\bf x}),\phi^a_j({\bf y}) \} =
     - 2 \ \varepsilon_{ijk} \ \phi^a_k({\bf x})
           \ \delta^3 ({\bf x} - {\bf y}) \ .}
                                                            \eqno(11)
$$
 All the constraints are of first class and the algebra is closed.
 The constraints $\varphi_k$ and $\phi^a_k$ generate the local-Lorentz
and new-type transformations in (6) respectively.
 In this canonical formulation, the on-shell reducibility (7) appears as
a linear dependence of the constraints,
$$
    D_a \phi^a_k - {\Lambda \over 6} \varphi_k = 0 \ .
                                                           \eqno(12)
$$
 Taking account of this relation, the number of independent constraints
is nine at each spatial point, which is just equal to the number of
canonical coordinates.
 Accordingly the physical degrees of freedom for this topological
case are $9 - 9 = 0$ as expected. \par

 On the other hand, in the Einstein gravity ($\alpha \not=$ 0),
we have to solve the constraint equations (3) which can be considered as
five linear equations for nine Lagrange multipliers $\Sigma _{a0}^k$
 [3].
 The solution is expressed with four arbitrary variables $N^a$
(shift vector) and $\utilde{N}$ (lapse density of weight $-1$),
$$
  \Sigma_{a0}^k = -{1 \over 4} \   \varepsilon_{abc} [ N^b B^c_k
       + \utilde{N}( B^b \times B^c )^k ] \ .
                                                          \eqno(13)
$$
 Substituting this result for the canonical action (9),
we now have four constraints, together with $\varphi_k$ in (10),
which are associated with the Lagrange multipliers $N^a$ and
$\utilde{N}$,
$$
\eqalign{ & C_a \equiv {1 \over 4} \  \varepsilon_{abc} B^b \cdot
                      \phi^c = B^b \cdot R_{ab} \approx 0 \ ,
\cr
          & C \equiv {1 \over 4} \  \varepsilon_{abc} ( B^a \times B^b)
              \cdot \phi ^c = ( B^a \times B^b ) \cdot
              ( R_{ab} - {\Lambda \over 24} \varepsilon_{abc}
                   B^c ) \approx 0 \ .}
                                                          \eqno(14)
$$
 Since the forms of these constraints are just the same as the ones in
the Ashtekar formalism [16,17], we can easily identify $\varphi_k, C_a$
and $C$ with the independent first class constraints corresponding to the
 generators of the local-Lorentz transformation, spatial diffeomorphism
and temporal diffeomorphism respectively.
 But the constraint algebra is now open as usual in the case of gravity.
 In this sense  the 2-form Einstein gravity is interpreted as a natural
covariantization of the Ashtekar formalism.
 The number of the physical degrees of freedom for the Einstein gravity
becomes $9 - 7 = 2$, the modes of the gravitational wave. \par

 As for the relation between the symmetry of the Einstein gravity and
that of its topological version, an important observation is
that all the constraints in the former are linear combinations of those
in the latter.
 Especially four diffeomorphism generators $C_a$ and $C$ in (14) are
linearly dependent on nine `new-type' generators $\phi^a_k$ in (10).
 This situation can be understood as follows; the large symmetry of the
topological model is partially broken in the Einstein gravity leaving
only local-Lorentz and diffeomorphism symmetries intact and as a result
the gravitational waves are induced.
 Note that the transition is caused by recovering the term with the
Lagrange multiplier field $\psi_{ABCD}$ in the action (1).
 With the help of the equations of motion derived from the action (1)
with $\alpha \not=0$, this $\psi_{ABCD}$ is determined to be
proportional to the anti-self-dual part of the Weyl (conformal) tensor
which just governs the modes of the gravitational wave.  \par

This leads us to an expectation that our topological model is a useful
tool to capture the global aspects of the 2-form Einstein gravity.
 We shall demonstrate an explicit example for such a use in the
following.
 In the Dirac approach for quantization [18], one has to impose
quantum conditions to choose physical wave functional $\Psi$.
 In the topological case with $\Lambda \not= 0,$  these conditions can
be expressed using the constraints (10) in $\omega_a^k$ representation,
$$
\eqalign{ & \varphi_k (\omega, \delta / \delta \omega) \Psi(\omega) =
            i D_a ( \delta / \delta \omega_a^k) \Psi(\omega) = 0  \ ,
\cr
          & \phi^a_k (\omega, \delta / \delta \omega) \Psi(\omega) =
            2 ( \varepsilon^{abc} R^k_{bc} + i {\Lambda \over 12} \,
            \delta / \delta \omega_a^k ) \Psi(\omega) = 0 \ .}
                                                             \eqno(15)
$$
 We can easily solve these equations to obtain the unique functional of
$\omega_a^k$,
$$
    \Psi(\omega) = \exp( {6i \over \Lambda} I_{C-S} ) \ ,
\qquad
    I_{C-S} \equiv \int d^3x \varepsilon^{abc} \omega_a \cdot
       ( \partial_b \omega_c + {2 \over 3} ( \omega_b \times
          \omega_c ) ) \ ,
                                                             \eqno(16)
$$
where $I_{C-S}$ is the Chern-Simons term on the three-dimensional
boundary.
 This type of solution is also found in a different version
of topological gravity [15].
 The functional $\Psi(\omega)$ can also be considered as the BRST
invariant `vacuum' because it is the unique representative annihilated
by the BRST operator defined using (17) below. \par

 Moreover $\Psi(\omega)$ becomes a special solution of all quantum
constraints including the Wheeler-DeWitt equation in the 2-form Einstein
gravity if the operator ordering is arranged as in (14).
 This ordering is adequate for our discussion since it is
consistent with the commutation relations among the constraint operators
[19].
 This $\Psi(\omega)$ is nothing but the Euclidean version of the wave
functional discovered by Kodama [20,21].
 Therefore the physical space of the topological model is contained
in that of the 2-form Einstein gravity. \par

 {}From this fact, we expect that more informations about the global
aspects of the Einstein gravity can be extracted by investigating its
topological version further.
 Henceforth we quantize the topological model with the method introduced
by Batalin, Fradkin and Vilkovisky [4-6], which seems to be one of the
most suitable procedures to examine the nature of topological quantum
field theory (TQFT).

\chapter{BRST quantization of topological 2-form gravity}%

 In the classification in [5,6], our system with the action (5)
corresponds to a first-stage reducible theory due to the presence of the
linearly dependent relation (12).
\footnote{{}^{\sharp4}}{In the following, we use the convention in [6]
and the notation,
       $F_{[\mu \nu]}
                      \equiv {1 \over 2}(F_{\mu \nu} - F_{\nu \mu})$. }
 Accordingly we have to enlarge the original phase space of
($ \omega_a^i , B^a_i $) by introducing (i) the Lagrange multipliers and
their conjugate momenta; bosonic fields ($\lambda^i , \pi_i$),
($\lambda_a^i , \pi^a_i$) and fermionic fields
($\lambda_1^i , \pi_{1i}$), (ii) the conjugate pairs of the fermionic
ghosts ($C^i , \bar{ {\cal P}}_i$), ($C_a^i , \bar{ {\cal P}}^a_i $),
($\bar{C}_i , {\cal P}^i $), ($\bar{C}^a_i , {\cal P}_a^i$) and (iii)
the conjugate pairs of the bosonic `ghosts for ghosts'
($C_1^i , \bar{{\cal P}}_{1i}$), ($\bar{C}_{1i} , {\cal P}_1^i$).
 The fields with subscript 1 belong to the class of the first-stage.
 Besides these fields, we need the conjugate pairs of the `extraghosts',
 bosonic (${}^{ex} \lambda^i ,{}^{ex} \pi_i$) and fermionic
(${}^{ex} \bar{C_i} , {}^{ex} {\cal P}^i$), in order to make the theory
covariant completely.
 The momenta for the Lagrange multipliers and ${}^{ex} \pi_i$ are
constrained to vanish.
 We assign to the ghosts $C^i, C_a^i$
(the anti-ghosts ${\bar C}_i, {\bar C}^a_i$) the ghost number $gh= 1
(-1)$ and $gh(\lambda_1^i,C_1^i,{\bar C}_{1i},{}^{ex} \lambda^i,{}^{ex}
\bar{C_i})=(1,2,-2,0,-1)$.
 The ghost number for each momentum field is set to be opposite to its
conjugate.
 Furthermore $\lambda_1^i ,{\bar C}_i ,{\bar C}^a_i ,
{}^{ex}{\bar C}_i , {\bar {\cal P}}_i$ and ${\bar {\cal P}}^a_i$ are pure
 imaginary while the other fields are real.\par

 Using these fields, the BRST charge is constructed as follows,
$$
 \eqalign { & \Omega = \int d^3x [C \cdot \varphi + C_a \cdot \phi^a
              + i C_1 \cdot ( D_a \bar{{\cal P}}^a - {\Lambda \over 6}
              \bar{{\cal P}} ) - ( C \times C ) \cdot \bar{{\cal P}}
\cr
            & - 2 ( C \times C_a ) \cdot \bar{{\cal P}}^a
              + 2 (C_1 \times C ) \cdot \bar{{\cal P}}_1
              +  \pi \cdot {\cal P}
              +   \pi^a \cdot {\cal P}_a   +  \pi_1 \cdot {\cal P}_1
              + {}^{ex} \pi \cdot {}^{ex} {\cal P} ] \ .}
                                                               \eqno(17)
$$
 It satisfies the nilpotency condition $\{\Omega,\ \Omega \}$ = 0
and is real, fermionic and $gh(\Omega) = 1$.
 If setting $\Lambda \not= 0$ and dropping the last extraghost term in
(17), this BRST charge becomes equivalent to the one derived in the TYMT
[10]. \par

 Next we have to choose the gauge fermion $\cal F$ to be pure imaginary
and $gh({\cal F})=-1$, in order to obtain the BRST invariant
(gauge fixed) action and the BRST transformation in covariant form,
$$
 \eqalign{
  {\cal F} = \int d^3x [
 & \bar{C} \cdot \chi + \bar{C}^a \cdot \chi_a + \bar{C}_1
   \cdot \chi_1   +  \bar{{\cal P}} \cdot \lambda
   + \bar{{\cal P}}^a \cdot \lambda_a  + \bar{{\cal P}}_1 \cdot \lambda_1
   + {}^{ex}\lambda \cdot \chi^\prime
\cr
          & + {}^{ex} \bar {C} \cdot \chi^{\prime  \prime}
            + i \beta \pi_1 \cdot (C_1 \times \bar{C}_1)
            - 2i \gamma (\bar{C}_1 \times C) \cdot
            (C_1 \times \bar{C}_1) ]  \ ,}                    \eqno(18)
$$
where
$$
\eqalign{
 & \chi_i \ \equiv - \, \partial_a \omega^a_i ,
\qquad \qquad \qquad \qquad
   \chi_a^i \equiv {1 \over 2} \, D_b ( {{\varepsilon ^b}_a}_c B^c_i )
        -2  \, ( \lambda \times \lambda_a )^i ,
\cr
 & \chi_{1i} \equiv i \, D_a C^a_i +
                  2 \, ( \lambda \times \lambda_1)^i ,
\qquad \,
   \chi^\prime_i \, \equiv \, D_a \bar C^a_i
                     +2 \, ( \lambda \times {}^{ex}\bar C )^i ,
\cr
 & \chi_i^{\prime \prime} \, \equiv \, D_a \lambda^a_i \ ,}
                                                            \eqno(19)
$$
and the parameters $\beta, \gamma$ are real.
 These choices turn out to be the gauge fixing conditions for
local-Lorentz, new-type and reducible symmetries respectively in the
final covariant expression,\footnote{{}^{\sharp5}}{We use the flat
background metric in the gauge fixing procedure for simplicity. }
$$
  \partial_\mu \omega^\mu_i = 0 \ ,
  \quad
  D_\nu \Sigma^{\mu \nu}_i = 0 \ ,
  \quad
  D_\mu C^\mu_i = D_\mu {\bar C}^\mu_i = 0 \ .
                                                          \eqno(20)
$$
 Then the BRST invariant (effective) action $S_{eff}$ is given by
$$
\eqalign{S_{eff} &=
         \int dt [\int d^3x ( \dot \omega_a \cdot B^a
                          + \dot \lambda \cdot \pi
                          + \dot \lambda_a  \cdot \pi^a
                          + \dot \lambda_1  \cdot \pi_1
                          + {}^{ex} \dot \lambda \cdot {}^{ex} \pi
                          + {\dot C} \cdot {\bar {\cal P}}
\cr
                        & + \dot C_a \cdot {\bar {\cal P}}^a
                          + \dot C_1 \cdot {\bar {\cal P}}_1
                          + \dot {\bar C} \cdot {\cal P}
                          + \dot {\bar C}^a \cdot {\cal P}_a
                          + \dot {\bar C}_1 \cdot {\cal P}_1
                          + {}^{ex} \dot {\bar C} \cdot {}^{ex}{\cal P})
                                             - H_{eff} ] \ .}
                                                             \eqno(21)
$$
 The original Hamiltonian $H_0$ vanishes due to the diffeomorphism
invariance of the theory and hence the effective Hamiltonian $H_{eff}$
is equal to $-\{{\cal F}, \Omega \}$.
 After excluding the conjugate momenta for the ghosts and the ghosts
for ghosts by using the equations of motion from (21),
identifying $ \lambda^i , \lambda_a^i , \lambda_1^i $
and $ {}^{ex} {\bar C_i} $ with  $ \omega_0^i , \Sigma_{a0}^i , iC_0^i$
and $ {\bar C}_0^i $, setting the parameters $ \beta = \gamma \,
(\equiv \kappa)$ and redefining the following fields,
$$
\eqalign{
 & \hat \pi^a_i \equiv \pi^a_i + 2 ( {\bar C^a} \times C )^i \ ,
  \qquad
  \hat \pi^0_i \equiv {}^{ex}\pi_i + 2 ({}^{ex}{\bar C} \times C )^i \ ,
\cr
 & \hat \pi_{1i} \equiv \pi_{1i} - 2 ( {\bar C_1} \times C )^i \ ,
  \qquad
  \tau^i \equiv {}^{ex}{\cal P}^i - 2 ( {}^{ex}\lambda \times C )^i \ ,}
                                                             \eqno(22)
$$
 we get
$$
     S_{eff} = S_{cl} + S_G + S_N + S_E  \ ,            \eqno(23)
$$
 where $S_{cl}$ is the starting action (5) and
$$
\eqalign{S_G &= \int d^4x [ \pi \cdot {\partial_\mu} {\omega^\mu}
         + {\partial^\mu} {\bar C} \cdot ( D_\mu C - {\Lambda \over 6}
           C_\mu )] \ ,
\cr
   S_N &= \int d^4x [ \hat \pi_\mu \cdot D_\nu \Sigma^{\mu\nu}
          + 2 D^\mu {\bar C^\nu} \cdot D_{[\mu} C_{\nu]} +
            {\Lambda \over 3} ({\bar C^\mu} \times C^\nu ) \cdot
            \Sigma_{\mu \nu}
\cr
   & -i {\hat \pi_1} \cdot D_\mu C^\mu
     + D^\mu {\bar C_1} \cdot D_\mu C_1
     + {\Lambda \over 3}\, i \, {\bar C_1} \cdot ( C^\mu \times C_\mu )
     - \tau \cdot D_\mu {\bar C^\mu}
\cr
   & - {}^{ex} \lambda \cdot ( D_\mu \hat \pi^\mu
     - {\Lambda \over 3}   ( {\bar C^\mu} \times C_\mu ) )
     - {i \over 2}  \ \varepsilon^{\mu \nu \rho \tau}
       ( D_\mu {\bar C_\nu} \times D_\rho {\bar C_\tau} ) \cdot C_1 ] \ ,
\cr
  S_E &=  \int d^4x [ - i \, \kappa ( \hat \pi_1 \times \hat \pi_1 )
  \cdot C_1 - {\Lambda \over 3}\, \kappa (C_1 \times {\bar C_1})^2 ] \ .}
                                                                \eqno(24)
$$
 All the terms in $S_N$ are obtained by the usual Faddeev-Popov procedure
 with the gauge fixing conditions (20) except for the last cubic ghost
term.
 This non-trivial term stems from the on-shell reducibility (7) of the
theory [13].  \par

 The BRST transformation in covariant form is given by
$$
 \eqalign{
 & \delta_B \omega_\mu^i = D_\mu C^i - {\Lambda \over 6} C_\mu^i \ ,
\cr
 & \delta_B \Sigma_{\mu\nu}^i = - 2 D_{[\mu} {C_{\nu]}}^i
       + 2 ( \Sigma_{\mu\nu} \times C )^i
       + i \, \varepsilon_{\mu\nu\rho\tau} ( D^\rho {\bar C}^\tau
       \times C_1)^i  \ ,
\cr
 & \delta_B C^i = ( C\times C )^i + {\Lambda \over 6} \, i \,  C_1^i \ ,
 \qquad \
   \delta_B {\bar C}_i = - \pi_i \ ,
 \qquad \qquad
   \delta_B \pi_i = 0 \ ,
\cr
 & \delta_B C_\mu^i = 2( C \times C_\mu)^i + i \, D_\mu C_1^i \ ,
   \ \ \ \delta_B C_1^i = - 2 ( C \times C_1 )^i \ ,
\cr
 & \delta_B {\bar C^\mu_i} = - \hat \pi^\mu_i
                         + 2 (\bar C^\mu \times C)^i \ ,
 \qquad
    \delta_B \hat \pi^\mu_i = 2 ( \hat \pi ^\mu \times C )^i
           + {\Lambda \over 3} \, i \, ( {\bar C^\mu} \times C_1)^i \ ,
\cr
 & \delta_B {\bar C}_{1i} = \hat \pi_{1i}
      + 2 ( {\bar C}_1 \times C )^i \ ,
 \qquad \ \
    \delta_B \hat \pi_{1i} = 2 ( \hat \pi_1 \times C )^i
           - {\Lambda \over 3} \,i \, ( {\bar C}_1 \times C_1 )^i \ ,
\cr
 & \delta_B {}^{ex}\lambda^i = \tau^i
       + 2 ( {}^{ex}\lambda \times C )^i \ ,
 \qquad \
   \delta_B \tau^i = 2 ( \tau \times C )^i
       -  {\Lambda \over 3} \,i \, ( {}^{ex}\lambda \times C_1)^i \ .}
                                                             \eqno(25)
$$
The off-shell nilpotency of the transformation is satisfied by
 all the fields except for the 2-form $\Sigma^i_{\mu\nu}$ and
$$
   {\delta_B^{\, 2}}\, \Sigma^i_{\mu\nu} =
     \, i \, \varepsilon_{\mu\nu\tau\rho} ( { \delta S_{eff} \over
          \delta \Sigma_{\rho\tau} } \times C_1 )^i \ ,
                                                              \eqno(26)
$$
which vanishes by the equations of motion.
 We have also derived all these results using the antifield formalism
introduced by Batalin and Vilkovisky [22]. \par

 If we choose the cosmological constant $\Lambda$ and the parameter
$\kappa$ in (24) to be zero and drop $S_G$ from (23) to keep the
local-Lorentz (SU(2)) symmetry, the effective action $S_{eff} = S_{cl} +
S_N$ becomes identical with the quantum action discussed in [13,14],
which describes a kind of `Schwarz-type' TQFT.   \par

 On the contrary, for the $\Lambda \not=0$ case, we expect that one can
reach the TYMT from our theory because of the equivalence of their
starting actions as noted previously.
 In fact, the quantum action of the TYMT appears if we change the gauge
fixing condition for the new-type symmetry as
$$
   D_{\nu} \Sigma^{\mu \nu}_i = 0 \
       \rightarrow \
                   D_{\nu} {^+ \Sigma^{\mu \nu}_i} =0  \ .    \eqno (27)
$$
 The superscript `+' means `self-dual' for the world indices
$\mu,\nu, \cdots$. Besides we redefine the following auxiliary
and ghost fields,
$$
  \pi^{\mu \nu}_i \equiv
              D^{[\mu} {\hat \pi}^{\nu]}_i
              - {\Lambda \over 3}(C^{[\mu} \times {\bar C}^{\nu]})^i \ ,
\qquad
  {\bar \chi}^{\mu \nu}_i \equiv D^{[\mu} {\bar C}^{\nu]}_i  \ .
                                                           \eqno (28)
$$
 Using the equations of motion derived from the variations with respect
to ${}^{ex}\lambda_i$, $\tau_i$ and $\Sigma^i_{\mu \nu}$, we get
$$
  S_{eff} = S_{cl}^{\prime} + S_G + S^{\prime}_N + S_E  \ ,   \eqno (29)
$$
 where $S_{cl}^{\prime}$ is the Pontryagin action (8) and
$$
\eqalign{ S^\prime_N = \int d^4x [\,
    &{6 \over \Lambda} {}^+\pi_{\rho \tau} \cdot {}^+ R^{\rho \tau}
    +{3 \over 4\Lambda} {}^+\pi^{\mu \nu} \cdot {}^+\pi_{\mu \nu}
    + 2 {}^+{\bar \chi}^{\mu \nu} \cdot D_\mu C_\nu
    - {\it i} {\hat \pi}_1 \cdot D^{\mu} C_{\mu}
\cr
   & + {\Lambda \over 3}{\it i} \bar C_1 \cdot (C^{\mu} \times C_{\mu})
     -{\it i} ({}^+{\bar \chi}_{\mu \nu}
       \times {}^+{\bar \chi}^{\mu \nu}) \cdot C_1
     + D^{\mu} \bar C_1 \cdot D_{\mu} C_1]  \ .  }    \eqno (30)
$$
 This $S_{eff}$ is equivalent to the action of the TYMT
[7-10].
 Also the elimination of $\Sigma_{\mu \nu}^i$ by using the equations of
motion brings about the off-shell nilpotent BRST transformation which is
the same as the one in the TYMT.
 We see that the inclusion of the cosmological term changes the
`Schwarz-type' TQFT into the `Witten-type' TQFT.

\chapter{Conclusion}%
 We have clarified the relation between the 2-form Einstein gravity
in four dimensions and its topological version, and performed the BRST
quantization for the TFT.
 With the benefit of several technical merits in the TFT, further
exploration of the theory will yield fruitful results for the Einstein
gravity and is also interesting from the mathematical point of view.
 Moreover, we can interpret our topological model as an unbroken phase
of the Einstein gravity.
 In fact, we have seen that the partial breakdown of the symmetry in
the TFT generates the Einstein gravity.
 It is sufficiently intriguing for us to pursue the mechanism explaining
this process as the spontaneous breakdown of the topological symmetry.

\vskip 2.5cm

\ack
We would like to thank Professors A. Hosoya and N. Sakai for valuable
discussion and reading the manuscript.
 We also acknowledge Drs. M. Abe and K. Amano for useful comments.



\REF\SEEW{E. Witten, Nucl.$\>$Phys.$\>$B311 (1988) 46.}

\REF\SEEW{E. Witten, Nucl.$\>$Phys.$\>$B323 (1989) 113.}

\REF\SEECDJM{R. Capovilla, J. Dell, T. Jacobson and L. Mason,  \nextline
Class.$\>$Quantum$\>$Grav.$\>$8 (1991) 41.}

\REF\SEEFVBV{E.S. Fradkin and G.A. Vilkovisky, Phys.$\>$Lett.$\>$B55
(1975) 224;
  \nextline  I.A. Batalin and G.A. Vilkovisky, Phys.$\>$Lett.$\>$B69
(1977) 309.}

\REF\SEEBF{I.A. Batalin and E.S. Fradkin, Phys.$\>$Lett.$\>$B122 (1982)
157.}

\REF\SEEH{M. Henneaux, Phys.$\>$Rep.$\>$26 (1985) 1.}

\REF\SEEW{E. Witten, Commun.$\>$Math.$\>$Phys.$\>$117 (1988) 353.}

\REF\SEEBS{L. Bauliew and I.M. Singer, Nucl.$\>$Phys.$\>$B (Proc. Suppl.) 5B
(1988) 12.}

\REF\SEELP{J.M.F. Labastida and M. Pernici, Phys.$\>$Lett.$\>$B212 (1988) 56.}

\REF\SEEIIKS{Y. Igarashi, H. Imai, S. Kitakado and H. So,
Phys.$\>$Lett.$\>$B227 (1989) 239.}

\REF\SEEPR{R. Penrose and W. Rindler, {\it `Spinors and Space-time'}
vol.1 (Cambridge University Press, 1984).}

\REF\SEEJS{T. Jacobson and L. Smolin, Class.$\>$Quantum$\>$Grav.$\>$5
(1988) 583.}

\REF\SEEBT{M. Blau and G. Thompson, Ann.$\>$of Phys.$\>$205 (1991) 130;
           \nextline  Phys.$\>$Lett.$\>$B255 (1991) 535.}

\REF\SEEBBRT{D. Birmingham, M. Blau, M. Rakowski and G. Thompson,
\nextline  Phys.$\>$Rep.$\>$209 (1991) 129.}

\REF\SEEH{G.T. Horowitz, Commun.$\>$Math.$\>$Phys.$\>$125 (1989) 417.}

\REF\SEEA{A. Ashtekar, Phys.$\>$Rev.$\>$D36 (1987) 1587; {\it `New
          Perspectives in Canonical Gravity'} (Bibliopolous, Naples,
          Italy, 1988).}

\REF\SEEART{A. Ashtekar, J.D. Romano and R.S. Tate, Phys.$\>$Rev.$\>$D40
(1989) 2572.}

\REF\SEED{P.A.M. Dirac, {\it `Lectures on Quantum Mechanics'}, Belfer
      Graduate School of Science (Yeshiva University, New York, 1964).}

\REF\SEEA{A. Ashtekar, Phys.$\>$Rev.$\>$Lett.$\>$57 (1986) 2244.}

\REF\SEEK{H. Kodama, Phys.$\>$Rev.$\>$D42 (1990) 2548.}

\REF\SEEI{H. Ikemori, in {\it `Proceeding of the Workshop on Quantum
             Gravity and Topology'}, edited by I. Oda (Institute for
             Nuclear Study, University of Tokyo, 1991). }

\REF\SEEBV{I.A. Batalin and G.A. Vilkovisky, Phys.$\>$Lett.$\>$B102
(1981) 27;  \nextline   Phys.$\>$Rev.$\>$D28 (1983) 2567.}

\null
\vfill

\refout
\vskip 2cm

\end